# The Relationship Between Smartphone Usage and Sleep Quality Amongst University Students

Hafsa Chaudhry, Hetvi Patel, Sai Teja Avadhootha, Sushanthik Reddy Poreddy, Swapan Gupta Chollati, Ujwala Namineni, and Huthaifa I. Ashqar

## Abstract

Gender differences were examined in sensitivity to sleep quality, in the context of blue light exposure from smartphones. Our hypothesis was created based on our journal article findings that females are more prone to be inclined to the prolonged usage of smartphones at bedtime and thus had less quality of sleep than males. The theory that usage affects sleep quality was due to the belief that the blue light emanating from the smartphone screen would disrupt our body's natural circadian rhythm, or sleep cycle, due to the blue light's ability to block a hormone called melatonin that controls and aids sleep. However, upon conducting regression tests and statistical analysis on our dataset, we found that our hypothesis was incorrect. Our dataset and analysis showed no relationship between smartphone usage and sleep quality in both males and females in young adults.

## Background and Aims

To explore the relationship between smartphone usage and sleep quality in young adults. Specifically, we will explore the function of the blue light filter from a smartphone and how it affects the quality and quantity of sleep in both university-aged males and females. We hypothesize that females are largely more vulnerable to smartphone usage at bedtime than males, therefore, are more prone to getting disrupted sleep quality than males. In contrast, we hypothesize that males were examined to be less susceptible to prolonged usage of smart devices prior to sleep, and would be recorded to have a better quality of sleep. Together, these findings would suggest that usage of smartphones negatively correlates to sleep quality.

## Literature Review

Research was conducted to find the link between sleep deprivation and smartphone usage in young adults and adolescents. Older adults differed in the findings due to their different schedules. Adolescents and young adults have the luxury to decide their use of electronic media without supervision constraints. These factors cause higher use of phones and media among young adults, such as university students (Demirci, 2015). It was found that adolescent girls have a greater tendency towards sleep quality problems. Smartphone dependence is associated with poor sleep quality among female college students (Wang, 2019).

Findings of the research articles have reported that ⅓ of our day is spent on our smartphones (Rayzah, 2021). This habit takes away the demands of other tasks, both physical and cognitive, causing deterioration of overall health (Rathakrishnan, 2021) and increases in mental health problems (Cui, 2020). Evidence supports the role of sleep in memory, attention, and learning as well as in the array of emotional regulations and behaviors (Kim, 2020). The impact of these findings shows negative effects on energy, sleep, physical and social activity, and academics (Saman, 2020) as well as anxiety, depression, and sleep disorders (Kaya, 2021). Studies have also found that overuse of cell phones exhibits increased abnormal behavior patterns such as staying up late to message or scroll the web (Saman, 2020) and state anxiety, trait anxiety, and depression were higher in the smartphone overuse group than in the normal use group (Demirci, 2015). The link between depression and insomnia created by smartphone usage at night was stated and examined in all 10 articles. Media content can create extreme excitement (Akçay, 2018) and the "fear of missing out", or FOMO, could drive this cycle (Belsare, 2020).

A common theme that emerged in the findings was regarding the relationship between melatonin and blue light. Blue light is a short wave light that delays the circadian clock phase, otherwise known as our sleep and wake cycle,  and suppresses the synthesis of melatonin (Rayzah, 2021). Melatonin is a natural hormone secreted by our pineal gland, which is inactive during the day and peaks during sleep. Reduction of the hormone makes it harder to fall and stay asleep. Making melatonin suppression the most likely contributing factor towards the relationship between phone and insomnia (Rayzah, 2021).

However, researchers have also found that improving health-related behaviors, such as nutritional behavior, stress management, physical activity, and development of a bedtime routine, have shown an enhancement towards sleep quality (Wang, 2019). By contrast, young adults with poor health-related behaviors, such as a tendency to eat unhealthy and high caloric diet, smoking and alcohol abuse, and self-harm, have shown a correlation towards poor sleep quality (Wang, 2019). The higher phone usage group may have a history of poor bedtime routines (Kim, 2020). It is not clear whether smartphones are responsible for anxiety and depression or the other way around (Belsare, 2020). Sleep quality is a mediator between technology after the onset of depression and anxiety, but smartphone use was not an independent predictor of sleep quality (Demirci, 2015). In addition, it is not likely that usage causes a substantial spike in mental health problems for most people (Belsare, 2020).  Specifically, female adolescents tend to have more disturbed sleep than males. This phenomenon starts from their first menstrual cycle, due to hormonal changes that affect and disturb sleep. Females tend to suffer insomnia, and irregular cycles and menstrual pain negatively impact sleep (Wang, 2019).

# Dataset

## Participants

We obtained this dataset from Kaggle.com where the dataset was originally published in India by Krupa Gajjar (Gajjar 2021). The dataset includes 46 participants identifying as females, males, and people unwilling to specify their genders aged 20 to 59 years old with a mean age of 24.76 years old. From the total number of participants, 26 are males, 18 are females, and two who preferred to not disclose their gender. Participants were asked questions such as their age, gender, hours of sleep time, hours of screen time on their smartphone, physical illness, sleep direction, number of meals consumed per day, whether they use a blue light filter or not, if they exercise physically, beverage preference (i.e. tea, coffee), smoke, or drink alcohol. Of the total participants, 91.3% do not smoke or drink alcohol, 50% sometimes physically exercise, 91% do not have a physical illness, and 54% use a blue light filter while using their smartphones at night. Inclusion criteria for questionnaire participants include possession of a smartphone, individuals with a normal sleep schedule (i.e. no night or grave shift), and general cognition. Exclusion criteria include not having a smartphone, being unable to record average hours of sleep per night obtained, and unwillingness to complete the entire questionnaire.

## Procedure

The dataset's questionnaire was focused on a younger population with a few outliers aged 39, 50, and 59 years old. The participants were first asked general questions such as their age, gender, and the number of meals consumed per day as well as any physical or mental illnesses that will affect their sleep quality and pattern. Next, participants were asked the number of hours they spent on their smartphone as excessive smartphone usage has been studied to reflect a shorter sleep duration and lower sleep quality. Regarding smartphone usage, participants were asked if they used a blue light filter while using their smartphones at night. When using a smartphone or any electronic device, the screen blocks our eyes to all colors. Exposure to all colors and light assists our body's circadian rhythm by keeping our light exposure consistent with nature. The color blue passes through our retinas without blockage, exposing our brain to light-turned images boosting our ability to stay alert and attentive (Harvard, 2020). While all light can block a person's ability to fall asleep by suppressing melatonin production, blue light tends to be stronger. This affects our body's ability to sleep more so than other colors (Harvard, 2020).

The questionnaire also asked the participants which direction they face while sleeping. Based on Ayurvedic medicine, the human head has a polar-like attraction and needs to face South to attract opposite poles while sleeping. Participants were asked if they exercised. Studies have shown that physical activity leads to a longer sleep cycle due to the reduction of stress and fatigue of the body. Participants were also asked if they smoke, drink alcohol, or have a preference for beverages such as tea or coffee, as they may have an impact on the participant's health that might lead to poor quality of sleep over time.

**Methods**

**Linear Regression**

For this project, the dataset obtained was used to run linear regression amongst:

1. Smartphone screen time usage and sleep duration across all genders

2. Smartphone screen time usage and sleep duration in females

3. Smartphone screen time usage and sleep duration in males

4. Age and sleep duration in females

5. Age and sleep duration in males

It was imperative to perform linear regression to test which factors have the most impact on sleep quality amongst the participants and estimate the properties of those factors. P-values are used to decide if the relationships between independent and dependent variables observed in this dataset also exist in the larger population outside of this project. The analysis will utilize p-values to test if independent variables such as age and gender affect sleep duration and whether to accept or reject our hypothesis.

P-values will be used to determine whether to accept or reject the null hypothesis depending on if p-values are less than our significance level, or alpha, of 5% (Frost 2021). Alpha measures the strength of evidence present, and if the p-value is less than alpha, we can reject the null hypothesis and prove the effect of the independent variable is statistically significant. By using p-values, we tested our hypothesis for relationships between sleep duration and screen usage duration in females and males, age and sleep duration in females and males, and sleep duration and screen usage duration across all genders and ages and for each of the variables mentioned.

**Correlation coefficients**

Coefficients from a linear regression model are used to determine if there is a negative or positive relationship between each of the independent and dependent variables. Coefficients are significant because they represent how much the mean of dependent variables changes if the independent variables change, while other variables are constant (Frost 2021). We used coefficients amongst sleep duration and screen usage duration in females and males, age and sleep duration in females and males, and sleep duration and screen usage duration across all genders and ages to determine if there was a relationship between the independent and dependent variables.

**R-squared and Residual Analysis**

R-squared is a measure that shows how well the model is fitted to the data. As R-squared values are measured on a scale of 0% to 100%, the higher the percentage, the better the model is fitted to the observations. Lower R-squared values indicate that the model is not able to explain the change in the dependent variable around its mean (Frost 2021). Residuals are important in regression analyses as they represent the difference between the observed (y) and predicted (y_predicted) variables. Ideally, the residual sum value should equal zero as that indicates that the model is unbiased; have equal amounts of values above and below zero. If the residual value is above zero, that indicates the model is underpredicting the values and is biased. If under zero, then the model is overpredicted, there aren't enough observed values, and there are too many predicted values. We performed R-squared and residual analysis amongst sleep duration and screen usage duration in females and males, age and sleep duration in females and males, and sleep duration and screen usage duration across all genders and ages.

**Descriptive Statistics**

Descriptive statistics are used to present and describe basic features of the data in a study as they provide straightforward summaries about the dataset and its measures. For this research project, we used the mean, and 25%, 50%, and 75% percentiles as our main analysis tools from descriptive statistics.

For this project, we used descriptive statistics to test the effect of the blue light filters while using smartphones on sleep duration for university-aged male and female students. We also used descriptive statistics to analyze whether sleep quality differs in males and females in terms of smartphone screen time usage.

**Results**

To test if sleep quality differs between young males and females, we set the independent variables in linear regression to be females, males, and age while the dependent variable was set to sleep duration. For descriptive statistical analysis, we tested the effect of smartphone screen time usage on sleep duration for females and males to see if there is a significant difference. We also used descriptive statistical analysis to analyze the effect of using a blue light filter while using a smartphone on females' and males' sleep duration and see if there was a significant change due to age.

**Smartphone screen time usage and sleep duration across all genders and ages**

First, we performed a statistical analysis on smartphone screen time usage and sleep duration across all genders and ages. After running the linear regression, we found the p-value to be 0.616, which is more than the alpha of 0.05. This indicates that we failed to reject the null hypothesis as smartphone screen time usage has no significant effect on sleep duration across all genders and ages. Next, we looked at the smartphone screen time coefficient across all genders and ages and the coefficient of -0.0801 shows that there is a negative relationship between smartphone screen time usage and sleep duration across all genders and ages. Third, we looked at the R-squared value of 0.006 which shows that there is little to no correlation between these two variables as the R-squared value is not as high as it should be.

The final analytical method we used is the residual analysis to conclude if our model is biased in the smartphone screen time usage and sleep duration across all genders and ages relationship. There is no pattern visible on the residual plot, but most residuals are negative. Thus indicating that the residual sum is negative, the model is overpredicted, and there are not enough observed values present.

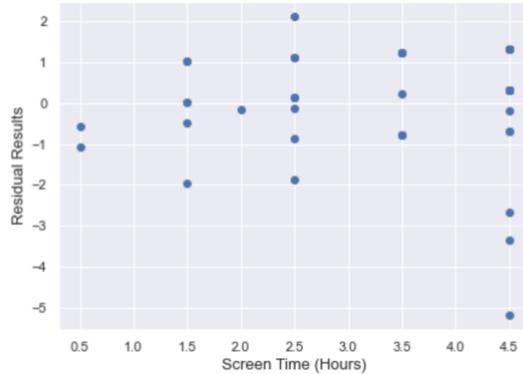

Figure 1.4. Residual Scatter Plot of Smartphone Screen Time vs Sleep Duration Across All Genders and Ages

**Smartphone screen time usage and sleep duration in females**

First, we performed a statistical analysis on smartphone screen time usage and sleep duration in females. After running the linear regression, we found the p-value to be 0.571 which is more than the alpha of 0.05. This indicates that we failed to reject the null hypothesis as smartphone screen time usage has no effect on sleep duration in females. Next, we looked at the smartphone screen time usage coefficient in females and the coefficient of 0.1169 shows that there is a positive relationship between smartphone screen time usage and sleep duration in females. Third, we looked at the R-squared value of 0.021 which shows that there is little to no correlation between these two variables as the R-squared value is too low.

The final analytical method we used is the residual analysis to conclude if our model is biased in the smartphone screen time usage and sleep duration in female relationships. We see from the residual plot that there is no pattern visible, but most residuals are positive. Thus indicating that the residual sum is positive, the model is underpredicted, and the model is biased.

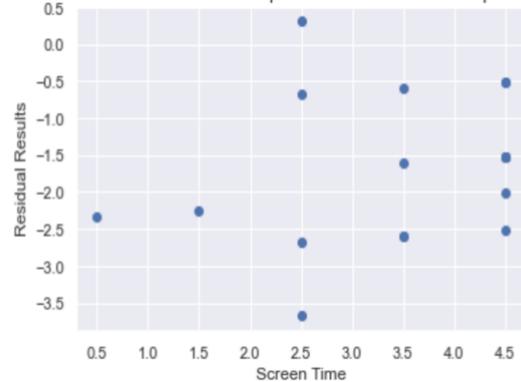

Figure 3.7. Residuls Scatter Plot of Smartphone Screen Time vs Sleep Duration in Females

**Smartphone screen time usage and sleep duration in males**

First, we performed a statistical analysis on smartphone screen time usage and sleep duration in males. After running the linear regression, we found the p-value to be 0.599 which is more than the alpha of 0.05. This indicates that we failed to reject the null hypothesis as smartphone screen time usage has no effect on sleep duration in males. Next, we looked at the smartphone screen time usage coefficient in males and the coefficient of -0.0954 shows that there is a negative relationship between smartphone screen time usage and sleep duration in males. Third, we looked at the R-squared value of 0.012 which shows that there is little to no correlation between these two variables as the R-squared value is too low.

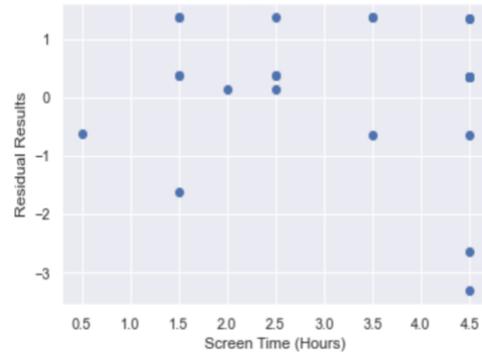

Figure 2.8. Residual Scatter Plot of Smartphone Screen Time vs Sleep Duration in Males

The final analytical method we used is the residual analysis to conclude if our model is biased in the smartphone screen time usage and sleep duration in male relationships. There is no pattern visible on the residual plot, but most residuals are negative. Thus indicating that the residual sum is negative, the model is overpredicted, and there are not enough observed values present.

**Age and sleep duration in females**

First, we performed a statistical analysis on age and sleep duration in females. After running the linear regression, we found the p-value to be 0.196 which is more than the alpha of 0.05. This indicates that we failed to reject the null hypothesis as age has no effect on sleep duration in females. Next, we looked at the age coefficient in females and the coefficient of -0.0808 shows that there is a negative relationship between age and sleep duration in females. Third, we looked at the R-squared value of 0.102 which shows that there is little to no correlation between these two variables as the R-squared value is too low.

The final analytical method we used is the residual analysis to conclude if our model is biased in the age and sleep duration in female relationship. The residual plot is left-skewed and all residuals are positive. This indicates that the residual sum is positive, the model is underpredicted, and the model is biased.

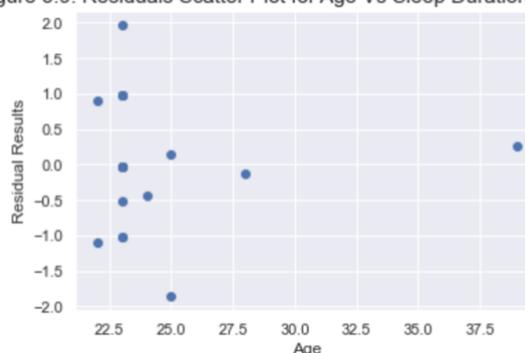

Figure 3.9. Residuals Scatter Plot for Age Vs Sleep Duration in Females

## Age and sleep duration in males

First, we performed a statistical analysis on age and sleep duration in males. After running the linear regression, we found the p-value to be 0.782 which is more than the alpha of 0.05. This indicates that we failed to reject the null hypothesis as age has no effect on sleep duration in males. Next, we looked at the age coefficient in males and the coefficient of 0.0077 shows that there is a positive relationship between age and sleep duration in males. Third, we looked at the R-squared value of 0.003 which shows that there is little to no correlation between these two variables as the R-squared value is too low.

The final analytical method we used is the residual analysis to conclude if our model is biased in the age and sleep duration in males relationship. We see from the residual plot that it is left-skewed, but almost all residuals are positive. This indicates that the residual sum is positive, the model is underpredicted, and there are not enough predicted values present.

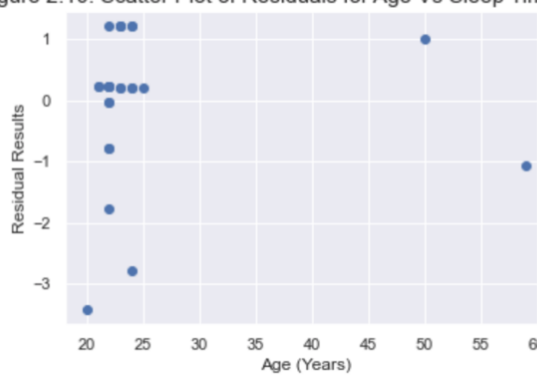

Figure 2.10. Scatter Plot of Residuals for Age Vs Sleep Time in Male

**Descriptive Statistics**

  The other analytical method that was used was descriptive statistical analysis. First, we used descriptive statistical analysis to note if a certain gender has a significantly different sleep duration due to the amount of smartphone screen time usage.  In Figure 7.3, the mean for female screen time is 3.444 hours and male screen time is 3.25 hours. But when compared to sleep duration, females slept an average of 6.916 hours, and males slept an average of 6.801 hours. Thus showing not a significant difference between the two genders in sleep duration due to smartphone screen time usage despite our literature review pointing out that females get lower sleep than males.

Figure 7.3

| | Age | | | | | | | | screen time | | | | | sleep time | | | | | | | |
|---|---|---|---|---|---|---|---|---|---|---|---|---|---|---|---|---|---|---|---|---|---|
| | count | mean | std | min | 25% | 50% | 75% | max | count | mean | ... | 75% | max | count | mean | std | min | 25% | 50% | 75% | max |
| **Gender** | | | | | | | | | | | | | | | | | | | | | |
| **Female** | 18.0 | 24.333333 | 3.910769 | 22.0 | 23.0 | 23.0 | 23.75 | 39.0 | 18.0 | 3.444444 | ... | 4.5 | 4.5 | 18.0 | 6.916667 | 0.988909 | 5.00 | 6.1250 | 7.00 | 7.750 | 9.0 |
| **Male** | 26.0 | 25.038462 | 8.838465 | 20.0 | 22.0 | 22.5 | 24.00 | 59.0 | 26.0 | 3.250000 | ... | 4.5 | 4.5 | 26.0 | 6.801731 | 1.192718 | 3.33 | 6.7575 | 7.00 | 7.750 | 8.0 |
| **Prefer not to say** | 2.0 | 25.000000 | 0.000000 | 25.0 | 25.0 | 25.0 | 25.00 | 25.0 | 2.0 | 4.500000 | ... | 4.5 | 4.5 | 2.0 | 4.750000 | 4.596194 | 1.50 | 3.1250 | 4.75 | 6.375 | 8.0 |

  Next, we used descriptive statistics to test if blue light filters usage while using smartphones results in higher sleep duration. In Figure 7.4, people who used a blue light filter averaged 25 years old and had a mean screen time usage of 3.28 hours and a mean of 6.66 hours of sleep. But people who did not use a blue light filter averaged 21 years old and had a mean screen time usage of 3.5 hours and a mean of 6.87 hours of sleep. Even though blue light filters block melatonin and affect the body's ability to fall asleep, the dataset here shows that the usage of a blue light filter while using a smartphone has no effect on the body's ability to fall asleep.

Figure 7.4

| | Age | | | | | | | | screen time | | | | | sleep time | | | | | | | |
|---|---|---|---|---|---|---|---|---|---|---|---|---|---|---|---|---|---|---|---|---|---|
| | count | mean | std | min | 25% | 50% | 75% | max | count | mean | ... | 75% | max | count | mean | std | min | 25% | 50% | 75% | max |
| filter | | | | | | | | | | | | | | | | | | | | | |
| no | 21.0 | 25.904762 | 9.684548 | 20.0 | 22.0 | 23.0 | 24.0 | 59.0 | 21.0 | 3.50 | ... | 4.5 | 4.5 | 21.0 | 6.872857 | 1.257566 | 3.33 | 6.0 | 7.0 | 8.0 | 9.0 |
| yes | 25.0 | 23.800000 | 3.488075 | 21.0 | 22.0 | 23.0 | 24.0 | 39.0 | 25.0 | 3.28 | ... | 4.5 | 4.5 | 25.0 | 6.660600 | 1.439586 | 1.50 | 6.5 | 7.0 | 7.0 | 8.0 |

   To compare if females or males had a longer sleep duration if they used a blue light filter

or not, we used a bar
graph to differentiate
the results. Figure 7.2
shows that females who
used the blue light filter
achieved almost seven
hours of sleep duration,
but females who did
not use a blue light
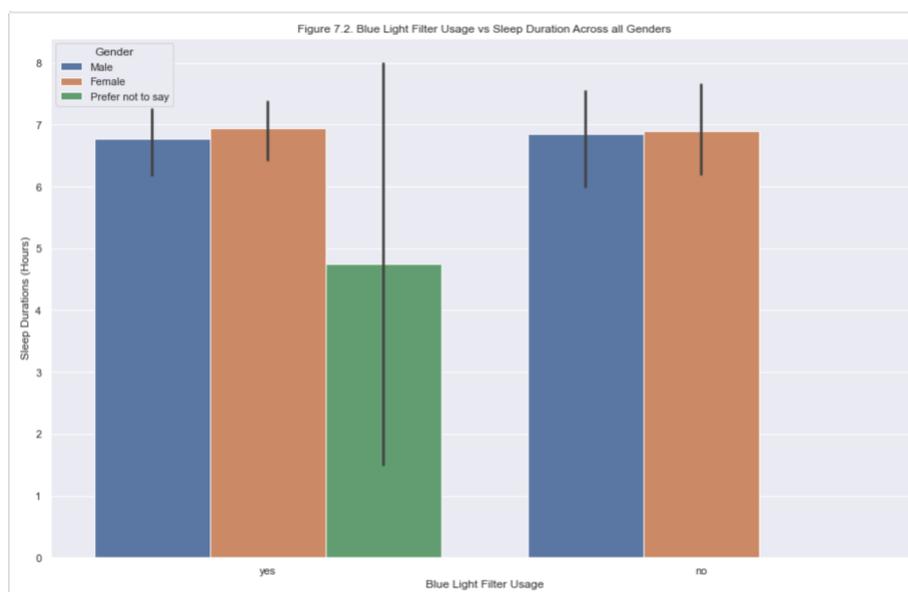
filter also got a sleep duration of almost seven hours. In the same figure, males who used a blue light filter got almost a little more than seven hours of sleep, and males who did not use the blue light got relatively the same amount of sleep as females who did not use the blue light. This figure shows that using a blue light filter while using a smartphone has no effect on females' or males' sleep duration.

## Analysis and Discussion

### The Process of Sleep

Sleep plays an important role in anyone's physical health. There are four phases of sleep, and the length of time spent in each sleep phase changes over a person's lifetime. The first phase of sleep is the shortest, lasting only a couple of minutes, and is when our brain undergoes the release of the alpha and theta waves. Those wave activities spike in the second phase. Upon entering the third phase, our brain produces delta waves, a slow wave that balms the body with restoration (Brady, 2019). As aforementioned, evidence supports the role of sleep in memory, attention, learning, emotional regulations, and behaviors (Kim, 2020). These qualities of energy and attentiveness are received in our third phase of sleep. The fourth and lightest slumber phase is REM sleep, or rapid eye movement sleep, and is where the majority of our dreams happen (Brady, 2019).

### Gender and Differences in Sleep

Gender plays a significant role in the modulation of sleep. In comparison to males, females naturally have an elevated slow delta wave-sleep in phase three, and a lower chance of wake-after-sleep onset. In other words, women have a lower chance of waking up suddenly in their sleep. While males have more difficulty maintaining sleep, sleep for shorter periods, and have more sleep-related breathing difficulties such as sleep apnea. Insomnia is one of the most prevalent sleep disorders, and its frequency rises as people age (Luca, 2015).

However, age is considered to be a deciding factor in sleep quality amongst all genders. Older women are more likely than older men to have poor sleep quality. Insomnia is one of the most prevalent sleep disorders, and its frequency rises as people age (Luca, 2015).

Based on our data, women are getting less sleep with age, but more smartphone screen time, while males are getting more sleep with age and less smartphone screen time. Though this discovery agrees with our hypothesis, our statistical analysis results revealed that smartphone dependencies were not significant predictors of sleep quality. Our statistical values were high, representing that a large difference exists between the two variables. Therefore our hypothesis is rejected and correlation does not imply causation. Just because a relationship exists between the two variables, does not mean one variable causes the other. These factors could be related to health issues. As aforementioned, starting from a female's first menstrual cycle, they tend to suffer from insomnia as irregular menstrual cycles and menstrual pain negatively impact sleep quality (Wang, 2019). Insomnia, as well as daytime sleepiness, is also caused by changes in hormones through the use of oral contraceptives for females. Oral contraceptives were released to the public in the 1960s, marking a method for the permanence of women in the workplace, by delayed pregnancies (Bezzera, 2018). Oral contraceptives work by blocking the hormonal release of gonadotropic in a women's brain, which stops the stimulation of hormones responsible for ovulation. This creates a chemical-induced, yet temporary, "menopause" (Hentschel, 2019). This process creates a shift in wave formations, and the short delta waves produced in phase three are lessened in production, while REM sleep in phase four is increased. In other words, a woman's body with oral contraceptives has a decreased sleep phase where cognitive and physical repairment is made, and an increased phase where her sleep is the lightest, and the easiest to be awoken (Hentschel, 2019). Other major health issues are perimenopause and menopause which disrupt quality sleep for females due to a decrease in the production of estrogen and progesterone. Menopause with age also results in hot flashes and sweat which can make it harder for an individual to sleep (Peters, 2020) as it disrupts our bodies natural methods of lowering body temperature to aid in sleep (Hentschel, 2019), and in return, results in poor sleep quality. Finally, menopause can result in depression, and depression is linked to insomnia (Peters, 2020).

**Conclusion**

The findings of this study reveal that there is no significant correlation between smartphone usage and sleep duration amongst the younger population of females and males. Smartphone usage includes screen time and blue light filter usage. Other factors such as age, high body mass index, obstructive sleep apnea, hypertension, menstrual cycles, consumption of large amounts of alcohol, caffeine, and nicotine, physical illness, and sleep environment are correlated to poor sleep quality for younger individuals than someone who leads an active and otherwise healthy lifestyle. Compared to males, females have longer slow-delta wave sleep and lower wake after sleep onset. However, women's sleep quality is negatively affected due to hormonal changes induced by oral contraceptives, irregular menstrual cycles, and menstrual pain. The study is different from the literature articles because of the limited dataset and unequal participants of the same gender, such as having 28 males and 16 females. Due to the smaller dataset, the regression model results were not accurate in reflecting the literature review we conducted. In terms of limitations, this study focused on a smaller group of participants with an unequal gender population and no indication of where the study was conducted and with what controls.